\documentclass[reprint,superscriptaddress,amsmath,amssymb,aps,prb,floatfix]{revtex4-2}
\usepackage{graphicx}
\usepackage{dcolumn}
\usepackage{amsmath, amssymb}
\usepackage{dsfont}
\usepackage{comment}
\usepackage{bm}
\usepackage{placeins}
\usepackage{xstring}
\usepackage[caption=false]{subfig}
\usepackage{color}
\usepackage[normalem]{ulem}
\usepackage[breaklinks=true]{hyperref}
\hypersetup{colorlinks=true,allcolors=blue}
\usepackage{cleveref}
\usepackage{xcolor}
\usepackage[normalem]{ulem}
\usepackage{placeins}
\usepackage{orcidlink}

\begin{document}

\preprint{APS/123-QED}

\title{Surface quantum response of plasmonic heterodimers:\\
revealing the hidden nonlocality}

\author{Ida Juliane Bundgaard\,\orcidlink{0009-0008-3424-0949}}
\email{Corresponding author: bundgaard@mci.sdu.dk}
\affiliation{POLIMA---Center for Polariton-driven Light--Matter Interactions,
University of Southern Denmark, Campusvej 55, DK-5230 Odense M, Denmark}

\author{Christos Mystilidis\,\orcidlink{0000-0002-8011-2154}}
\affiliation{POLIMA---Center for Polariton-driven Light--Matter Interactions,
University of Southern Denmark, Campusvej 55, DK-5230 Odense M, Denmark}

\author{Frederik~Wulff~Christensen\,\orcidlink{0009-0000-3950-4538}}
\affiliation{POLIMA---Center for Polariton-driven Light--Matter Interactions,
University of Southern Denmark, Campusvej 55, DK-5230 Odense M, Denmark}

\author{Lorenz~Huber\,\orcidlink{0009-0000-8321-4896}}
\affiliation{Institute of Physics, University of Graz,
Universit\"{a}tsplatz 5, A-8010 Graz, Austria}

\author{P.~Elli~Stamatopoulou\,\orcidlink{0000-0001-9121-911X}}
\affiliation{Institute of Nanotechnology, Karlsruhe Institute of Technology,
Kaiserstr. 12, DE-76131, Germany}

\author{Carsten Rockstuhl\,\orcidlink{0000-0002-5868-0526}}
\affiliation{Institute of Nanotechnology, Karlsruhe Institute of Technology,
Kaiserstr. 12, DE-76131, Germany}
\affiliation{Institute of Theoretical Solid State Physics, Karlsruhe
Institute of Technology, Kaiserstr. 12, DE-76131, Germany}

\author{P.~A.~D.~Gon\c{c}alves\,\orcidlink{0000-0001-8518-3886}}
\affiliation{POLIMA---Center for Polariton-driven Light--Matter Interactions,
University of Southern Denmark, Campusvej 55, DK-5230 Odense M, Denmark}

\author{Ulrich~Hohenester\,\orcidlink{0000-0001-8929-2086}}
\affiliation{Institute of Physics, University of Graz,
Universit\"{a}tsplatz 5, A-8010 Graz, Austria}

\author{N.~Asger~Mortensen\,\orcidlink{0000-0001-7936-6264}}
\affiliation{POLIMA---Center for Polariton-driven Light--Matter Interactions,
University of Southern Denmark, Campusvej 55, DK-5230 Odense M, Denmark}
\affiliation{D-IAS---Danish Institute for Advanced Study,
Campusvej 55, University of Southern Denmark, DK-5230 Odense M, Denmark}

\author{Christos Tserkezis\,\orcidlink{0000-0002-2075-9036}}
\email{Corresponding author: ct@mci.sdu.dk}
\affiliation{POLIMA---Center for Polariton-driven Light--Matter Interactions,
University of Southern Denmark, Campusvej 55, DK-5230 Odense M, Denmark}
\affiliation{D-IAS---Danish Institute for Advanced Study,
Campusvej 55, University of Southern Denmark, DK-5230 Odense M, Denmark}

\begin{abstract}
We probe the quantum response of plasmonic heterodimers
with extended (plane waves) and localized (point dipoles
and electron beams) sources. Implementing the
surface-response formalism based on Feibelman parameters,
we explore nanosphere heterodimers where one particle is
characterized by electron spill-in, while the other by
spill-out. We identify situations where the two quantum
surface corrections that act in opposite spectral-shift
directions can cancel out, producing mesoscopic spectra
that do not differ from the predictions of the
local-response approximation. This picture changes
dramatically once localized sources are implemented,
allowing one to selectively probe the quantum response
of the particles independently. Such probing occurs in
terms of higher-order modes in the case of electron beams,
or for the complete modal landscape in the case of point
dipole sources. This description offers a new insight
into the mechanisms governing light--matter interactions
in extreme plasmonic architectures, and opens new
perspectives for the understanding and tailoring of
plasmonic--quantum emitter architectures.
\end{abstract}

\maketitle

\section{Introduction}\label{sec:Introduction}

Mesoscopic or quantum plasmonics has grown into an
individual research sub-area in the past 15 years, as a
result of a continuous effort to better understand the
microscopic quantum mechanisms governing the interaction
of light with metallic nanostructures characterized by
few-to-sub-nm geometrical
details~\cite{mortensen_nanoph10a,zhu_natcom7,stamatopoulou_omex12}.
With plasmonics gradually departing from classical
electrodynamics and entering the regime of fundamental
condensed-matter physics, several size-related effects
acquire non-negligible importance. These include the
quantum pressure of the free-electron
gas~\cite{ruppin_prl31,abajo_jpcc112,raza_jpcm27} (rooted
in Pauli's exclusion principle), size-dependent resonance
broadening due to the reduced mean-electron
path~\cite{kreibig_zphys224,kreibig_surfsci156} or due
to increased surface-enabled Landau
damping~\cite{khurgin_acsphot4,Uskov_plasmonics9,tserkezis_ijmpb7,li_njp15},
and electron spill-out in low-work-function
metals~\cite{zuloaga_nl9,teperik_prl110,stella_jpcc117}.
In plasmonic cavities, the latter effect can even lead
to tunneling through extreme plasmonic
nanogaps~\cite{esteban_natcom3,ding_adpr4}. When one
operates down at the extreme nanoscale, all the
aforementioned effects become not just curiosities, but
factors that drastically influence the near- and far-field
properties of the system.

Among the different manifestations of mesoscopic effects
and the applications that they can affect, of particular
interest is their role on modifying the strength of
light--matter interactions~\cite{monticone_omex15}. Quantum
emitters (QEs) placed in the proximity of plasmonic
nanoparticles (NPs) or nanocavities experience a
dramatically altered electromagnetic (EM)
environment~\cite{anger_prl96,koenderink_ol35},
opening exciting possibilities for fluorescence enhancement or
inhibition~\cite{drexhage_bbpcl72,xu_prl93,fu_lpr3,su_acsabm4}, 
single-photon
generation~\cite{chang_prl97,koenderink_nl9,kumar_acsphot8},
and quantum applications such as
entanglement~\cite{martin_prb84,fernandez_acsphot5}. The role
of mesoscopic effects on the performance of plasmonic systems
as QE cavities has been explored in the last decade, in both the
weak~\cite{vielma_jcp126,tserkezis_nscale8,jurga_jpcc121} and
the strong-coupling
regimes~\cite{tserkezis_acsphot5a,ciraci_nanoph8,bundgaard_josab41}.
A common observation within all such works seems to be that
far-field observables quantifying the QE-plasmon coupling might
experience frequency shifts inheriting the response of the
corresponding plasmonic modes, but near-field observables are
generally more sensitive to quantum corrections in the
response of the metal.

In the zoo of plasmonic architectures that have been
introduced over the years, dimers of closely arranged
nanospheres or NPs of other simple shapes play a key role,
as they constitute simple and efficient plasmonic
cavities that provide field enhancement and confinement between NPs~\cite{romero_oex14,toscano_oex20,zuloaga_nl9,acimovic_nn3}.
In the case of sphere dimers, intuitive understanding of
the optical response can be drawn from the analytically
solvable single-NP case~\cite{mie_annphys330} (alternatively,
in the simpler picture of the quasistatic regime) in
combination with the concept of plasmon
hybridization~\cite{prodan_sci302}. Further, the inherent
symmetries of such architectures render them particularly
attractive from a computational point of view, offering
significant acceleration in established numerical
techniques~\cite{Ciraci2013} or allowing for the design of
ad hoc and semi-analytical ones~\cite{Beutel2024,Zheng2026ieee}.
The aforementioned insights and the ease of computation make
(spherical) plasmonic nanodimers a theorist's workhorse.
Dimers of identical NPs are often the first obvious
candidates for any study that focuses on understanding an
effect or displaying a new
functionality~\cite{nordlander_nl4,chandra_jacs134}. In the
context of light--matter interactions, nanosphere dimers
have been considered for, e.g., fluorescence
enhancement~\cite{tserkezis_prb96},
Raman scattering~\cite{li_admat36}, and surface-plasmon
amplification by stimulated emission of
radiation~\cite{warnakula_prb100}. One straightforward route
to slightly increase complexity and thus produce a richer
optical response is offered by heterodimers, where the two
spheres can either have different sizes or be composed of
different materials~\cite{sheikholeslami_nl10,cha_bkcs36}.
The role of heterodimers as QE environments is much less
investigated~\cite{huang_jpcc130}, and the influence of
mesoscopic effects is practically missing.

Here, taking a step to fill this gap, we study the optical
response of sphere heterodimers, within the context of the
surface-response formalism (SRF) that incorporates all
mesoscopic corrections into the Feibelman $d$ parameters.
We focus on a system in which one sphere is characterized
by electron spill-in, while the other exhibits spill-out,
and pose the question whether these two effects can
effectively cancel out, such that the resulting spectra
might behave as if no quantum correction is applied, in terms
of both the energy and the linewidth of the resonances.
In a perturbative picture, the two surface-response functions
would contribute with equal strength but opposite signs in
an otherwise symmetric dimer, thereby canceling each other.
Motivated by this insight, we see that, under certain
conditions, the far-field spectra of the system within a
mesoscopic description can appear practically indistinguishable
from those obtained within the local-response approximation
(LRA). We demonstrate the effect through a toy-model
calculation and then identify realistic material and size
parameters for which this can indeed happen. We further
proceed to explore the system with electron-beam or
dipole excitations, and explain when and how a localized
source can in fact reveal the nonlocality masked in the
response of the plane-wave excitation. With localized
sources describing the local density of optical states
(LDOS) much better, and accurately representing QEs in realistic
applications, we anticipate that our findings will provide
new insights into tailoring light--matter interactions with
extreme plasmonic nanocavities.

\section{Methods}\label{Sec:Methods}

\begin{figure*}[ht]
\centering\includegraphics[width=0.8\linewidth]{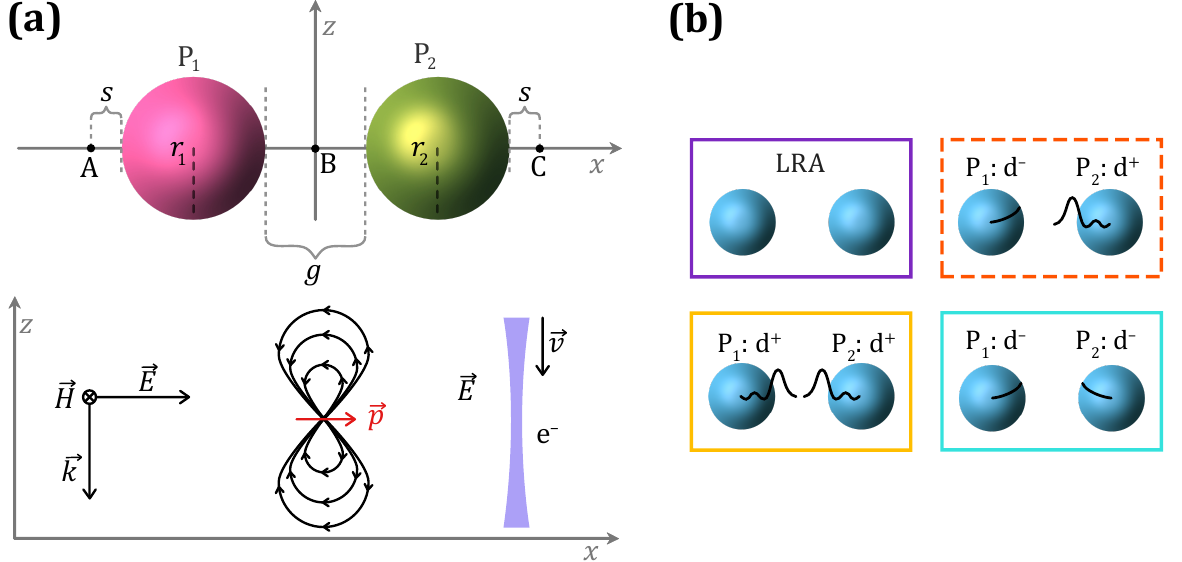}
\caption{(a) Sketch of two plasmonic nanospheres of different
materials and possibly different radii $r_{1}$ and $r_{2}$,
separated by a surface-to-surface gap of width $g$. The dimer
can be excited by a plane wave polarized along the dimer axis
(taken as $x$), or by a point dipole of dipole moment 
$\vec{p} \parallel x$, or by a swift electron beam traveling
along $z$ with velocity $\vec{v}$, as sketched in the lower
part of the panel. Positions A, B, and C correspond to the
dipole placements (at distance $s$ from the surface of the
nearest NP, in the case of positions A and C), or where the
electron beam passes by.
(b) The four different dimer types.
Top-left: the classical LRA approach.
Top-right: the left NP is characterized by spill-in ($d^{-}$),
while the right one by spill-out ($d^{+}$), as implied by the
sketches of induced-charge profiles.
Bottom-left: both NPs are characterized by spill-out.
Bottom-right: both NPs are characterized by spill-in.
In the four sketches, the colored boxes follow the line-color
code used in the remainder of the paper.
}\label{fig:sketch}
\end{figure*}

In what follows, we use the boundary-element method
(BEM) in the open-source implementation 
\textsc{nanobem}~\cite{hohenester_comphy276,Hohenester2024}
to study the response of heterodimers composed of plasmonic
spheres, as sketched in Fig.~\ref{fig:sketch}(a). The two
spheres are made of different plasmonic materials. Their
radii $r_{i}$ can range from $4$ to $20$\,nm, while their
separation is kept between $g = 1$\,nm and $g = 5$\,nm. For
such parameters, mesoscopic effects can be non-negligible
both due to the small NP size and, more importantly, due
to the decreasing plasmonic nanogaps, where the near-field 
enhancement can be strongly reduced in comparison to LRA.
To describe mesoscopic effects, we resort to SRF, where
nonlocality is encoded into a set of surface polarization
and current corrections, as captured by the Feibelman
parameters for metal--dielectric interfaces~\cite{feibelman_pss12}.
Assuming a \emph{planar} interface with translational
invariance in the $yz$ plane and with $x = 0$ the plane
delimiting the metal--dielectric interface, Feibelman
parameters in the long-wavelength, local 
limit~\cite{babaze_nanoph12} are defined as 
\begin{align}\label{Eq:d_parameters}
d_{\perp} (\omega) =
\frac{\int \mathrm{d} x\; x\, \rho (x, \omega)}
{\int \mathrm{d} x\; \rho (x, \omega)}\,,
\quad \quad
d_{\parallel} (\omega) =
\frac{\int \mathrm{d} x\; x\ \partial_{x} J_{y} (x, \omega)}
{\int \mathrm{d} x\; \partial_{x} J_{y} (x, \omega)}\,,
\end{align}
with $d_{\alpha} (\omega)$ 
($\alpha = \perp, \parallel$) being, in general, complex-valued
and dispersive (in accordance with the principles of causality).
Here, $\rho(x, \omega)$ is the induced charge density,
$J_{y}(x, \omega)$ is the induced current density in the
direction parallel to the metal--dielectric interface, while
$\omega$ is the angular frequency. As it becomes obvious from
their definition, the Feibelman parameters have dimensions of
length. The real part of the perpendicular Feibelman parameter,
$d_{\perp}$, expresses the shift of the centroid of the
induced charge with respect to the assumed interface, and it
captures electron spill-out ($d_{\perp} > 0$) in simple metals
like alkalis, or spill-in ($d_{\perp} < 0$) due to enhanced
screening (e.g., due to d-band electrons) in metals like the
noble ones. Similarly, the real part of $d_{\parallel}$
expresses the shift of the normal derivative of the induced
surface current. In what follows, we take $d_{\parallel}(\omega) = 0$,
as appropriate for charge-neutral metals~\cite{Apell1982}.
In both cases, the imaginary parts introduce losses related
to surface scattering and surface-enabled Landau damping.
SRF has been implemented in 
\textsc{nanobem}~\cite{hohenester_prb105,huber_jpcc129}, allowing
for efficient numerical evaluation of nonlocal effects in NPs
or NP aggregates of any shape, potentially supported by a
substrate~\cite{Hohenester2024}.
It is important to note that the Feibelman parameters, as
introduced above, correspond to a semi-infinite flat
metal--dielectric interface which, in extreme cases, might call
for additional corrections, either in the form of geometrical
factors~\cite{christensen_prl118} or with the full adoption
of a nonlocal framework~\cite{babaze_nanoph12,huber_jpcc129}.
The latter, however, relies on nonlocal parameters obtained
from \emph{ab initio} calculations for all parallel wave vector
components for every different metal--dielectric interface,
a computationally formidable task that goes beyond the scope
of this more illustrative work.

In the examples explored here, we first consider a toy model,
where the heterodimer is composed of Drude metals, with
relative permittivity given by
\begin{align}\label{Eq:Drude}
\varepsilon_{\mathrm{m}}(\omega) = 
\varepsilon_{\mathrm{b} }(\omega) - 
\frac{\omega_{\mathrm{p}}^{2}}
{\omega^{2} + \mathrm{i} \omega \gamma}\,.
\end{align}
with the same plasma energy $\hbar \omega_{\mathrm{p}} = 4$\,eV,
and the same damping parameter $\hbar \gamma = 0.1$\,eV.
In Eq.~(\ref{Eq:Drude}), $\varepsilon_{\mathrm{b}}(\omega)$
is a background permittivity to potentially account for
mechanisms such as interband transitions. In the toy model,
this is equal to 1. The $d_{\perp}(\omega)$ parameters of
the two NPs are assumed to be constant and opposite,
$d_{\perp} = \pm 0.4$\,nm, i.e., the real part inspired by
Ref.~\cite{goncalves_nl23}, meaning that one NP shall exhibit
spill-in while the other shall be characterized by an equal
amount of spill-out. Subsequently, realistic materials are
considered: sodium (Na) for the spill-out case, with
$\hbar \omega_{\mathrm{p}} = 5.89$\,eV and
$\hbar \gamma = 0.1$\,eV; and silver (Ag), described by the
experimental data of Johnson and Christy~\cite{johnson_prb6},
interpolated in the spectral window of interest herein. For
the Feibelman $d_\perp(\omega)$-parameter of Na, we employ
the density-functional-theory (DFT)-data from
Refs.~\cite{yan_prl115,christensen_prl118}, as fitted with a
series of Lorentzians in Ref.~\cite{goncalves_natcom11}. On
the other hand, for Ag we derive the values of $d_\perp(\omega)$
from the specular reflection model (SRM) in the long-wavelength
limit,
namely~\cite{ford_physrep113,feibelman_pss12,eriksen_nanoph13}
\begin{align}\label{Eq:d_SRM_model}
d_{\perp}^{\text{SRM}}(\omega) = 
-\frac{2}{\pi} 
\frac{\varepsilon_{\mathrm{m}}(\omega) 
\varepsilon_{\mathrm{d}}}
{\varepsilon_{\mathrm{m}}(\omega) - 
\varepsilon_{\mathrm{d}}} 
\int_{0}^{\infty} 
\frac{\mathrm{d} k_{\mathrm{L}}}{k_{\mathrm{L}}^{2}} 
\left[\frac{1}
{\varepsilon_{\mathrm{L}} (k_{\mathrm{L}}, \omega)} - 
\frac{1}{\varepsilon_{\mathrm{m}}(\omega)}\right]\,,
\end{align}
where $\varepsilon_{\mathrm{m}}$ is the bulk permittivity of
the metal, $\varepsilon_{\mathrm{d}}$ the one of the dielectric
environment, and $\varepsilon_{\mathrm{L}}$ is the longitudinal
permittivity of the metal within HDM, namely 
\begin{align}\label{Eq:HDMepsL}
\varepsilon_{\mathrm{L}}(k_{\mathrm{L}}, \omega) = 
\varepsilon_{\mathrm{b} }(\omega) - 
\frac{\omega_{\mathrm{p}}^{2}}
{\omega^{2} + \mathrm{i} \omega \gamma - 
\beta^{2} k_{\mathrm{L}}^{2}}\,.
\end{align}
Here, $k_{\mathrm{L}}$ is the longitudinal wave number in the
metal~\cite{raza_jpcm27} (obtained from the dispersion relation
$\varepsilon_{\mathrm{L}}(k_{\mathrm{L}}, \omega)$ = 0, with
the transverse one being 
$k = \frac{\omega}{c} \sqrt{\varepsilon_{\mathrm{m}}}$ and
$c$ being the speed of light in vacuum), and
$\beta \propto v_{\mathrm{F}}$ is the hydrodynamic parameter,
dependent on the Fermi velocity $v_{\mathrm{F}}$ of the
metal~\cite{ciraci_sci337}. In the high (optical) frequency
limit, 
$\beta^{2} = 3v_{\mathrm{F}}^{2}/5$~\cite{halevi_prb51,wegner_prb107}.
Finally, the background permittivity $\varepsilon_{\mathrm{b}}$
is obtained by subtracting a Drude-model fit with
$\hbar \omega_{\mathrm{p}} = 9.17$\,eV and 
$\hbar \gamma = 0.02$\,eV from the experimental bulk data.
Given such expressions, it is possible to perform the integration
of Eq.~\eqref{Eq:d_SRM_model}, leading to\footnote{Despite the
simplicity, note that this expression involves the square root
of a complex function, and $2\pi$ discontinuities must be
considered to ensure a continuous function.}
\begin{equation}\label{Eq:d_HDM}
d_{\perp}^{\mathrm{SRM}}(\omega) = 
\mathrm{i} 
\frac{\varepsilon_{\mathrm{m}}( \omega) 
\varepsilon_{\mathrm{d}}}
{\varepsilon_{\mathrm{m}}(\omega) - 
\varepsilon_{\mathrm{d}}}
\frac{\beta}{\omega_{\mathrm{p}} 
\sqrt{\varepsilon_{\mathrm{b}}(\omega)}} 
\left( \frac{\varepsilon_{\mathrm{b}}(\omega)}
{\varepsilon_{\mathrm{m}}(\omega)} - 1 \right)^{3/2}\,, 
\end{equation}
which is the expression we use in \textsc{nanobem}. The in-tandem
use of the DFT-level parameters for Na and HDM-level parameters
for Ag may appear at first inconsistent since there is a
significant mismatch in the accuracy level of the two
approximations, with HDM being a first step towards more mature
iterations of the (already approximate) orbital-free
DFT~\cite{toscano_natcom6,yan_prb91,ciraci_prb93}. This, however,
is a pragmatic choice, reflecting the relative scarcity of
tabulated Feibelman parameters or models thereof for different
materials~\cite{yang_nat576,mortensen_nanoph10b,chen_pnas122},
On the other hand, SRM has been shown to be reasonably accurate
in comparison with granular random-phase approximation (RPA)
calculations for Ag~\cite{rodriguez_optica8}, capturing
all its salient characteristics. As will become
evident in what follows, there is no reason to doubt that
for a different set of Feibelman parameters different geometry
choices will still lead to the cancellation under investigation.
The exact dispersion of both considered values for
$d_{\perp}(\omega)$ is plotted when discussing the Na--Ag
example in Fig.~\ref{fig:AgNaMap}(a). We note that the
$d_{\perp}(\omega)$ parameter of Ag exhibits a negative
imaginary part which, at a first glance, might appear to be
violating passivity (locally; but not globally). Nevertheless,
this negative imaginary part is what complies with Kramers--Kronig
relations for the given real part that exhibits a Lorentzian
antiresonance form to capture spill-in~\cite{svendsen_jpcm32}.

Validation of the results pertaining to plane-wave and dipole
excitations and realistic materials was performed by the
open-source software
\textsc{openmustanc}~\cite{zheng_arxiv2026}, which is a
multiple-scattering code relying on $\mathbf{S}$-matrix formalism.
For the case of plane waves, we observed excellent agreement
between the two independent approaches. For dipoles (and especially
for the examples below where the dipole-to-surface distance is
rather small), we observed challenging convergence yet
satisfactory agreement between \textsc{nanobem} and
\textsc{openmustanc}. In terms of the spectral positions of
the plasmonic resonances, the disagreement does not exceed
$0.4$\%, and therefore differences concern solely the amplitudes
thereof. It is interesting to note that the geometric perspective
we discuss herein is strongly reflected in the numerics. The convergence
for the case of a dipole oscillating to the right of the Na
nanosphere is very challenging, and can be intuitively attributed
to the effectively ---due to spill-out--- smaller gap. For
transparency, we report that the discretization (using in all
cases the native engine of \textsc{nanobem}) consists of
$1,564$ triangular elements per sphere for the plane-wave
case, while it is refined to $1,796$ triangular elements
per sphere for the dipole excitation.

\section{A toy model for spill-in/out overlap}

\begin{figure*}[ht]
\centering\includegraphics[width=0.8\linewidth]{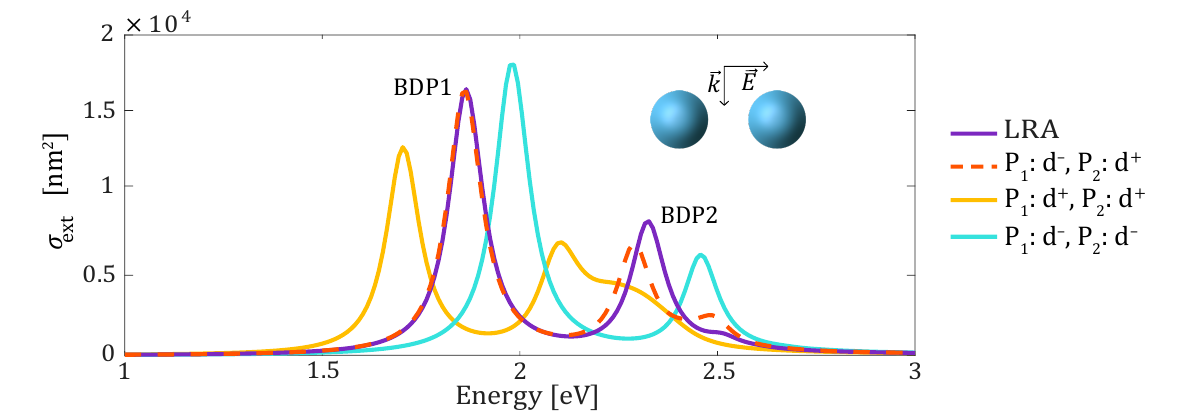}
\caption{Extinction spectra for a nanosphere dimer of
two otherwise identical NPs ($r_{1} = r_{2} = 16$\,nm,
Drude metals with $\hbar \omega_{\mathrm{p}} = 4$\,eV,
and $\hbar \gamma = 0.1$\,eV) separated by a gap distance
of $g = 3$\,nm, described by different combinations of
Feibelman parameters. The purple line shows the classical
LRA prediction, while the dashed orange considers two
opposite Feibelman parameters $d_{\perp} = \pm 0.4$\,nm.
The yellow spectrum assumes both NPs having
$d_{\perp} = + 0.4$\,nm, while the light blue one 
corresponds to both having $d_{\perp} = -0.4$\,nm.
In all cases, the dimer is excited by a plane wave polarized
along the dimer axis, as shown in the inset.
}\label{fig:toymodelPW}
\end{figure*}

The main purpose of this work is to explore whether 
spill-in and spill-out can effectively cancel each other
out in a dimer. To demonstrate this, we begin with a
basic toy model: two spheres of equal radii, here
chosen as $r_{1} = r_{2} = 16$\,nm, with permittivities
given by the local Drude model described in 
Sec.~\ref{Sec:Methods} [Eq.~(\ref{Eq:Drude})] and opposite
Feibelman parameters $d_{\perp} = \pm 0.4$\,nm, are placed
at a surface-to-surface distance $g = 3$\,nm. The dimer is
irradiated by a plane wave polarized along the dimer axis,
as sketched in the inset of Fig.~\ref{fig:toymodelPW}.

Figure~\ref{fig:toymodelPW} shows the extinction spectrum
of the dimer, as calculated within LRA (purple line)
and within SRF (orange dashed line). For comparison, we
also show the corresponding spectra if both NPs were
characterized by spill-out (yellow line) or spill-in
(light blue line). It is evident from the figure
that, for the given parameter set, the resonance and 
linewidth of the main bonding dimer plasmon (BDP1) at
$\sim 1.86$\,eV practically coincide within LRA and SRF.
In fact, even for the second-order BDP mode (BDP2) at
$\sim 2.3$\,eV, the two resonances are quite close
in the calculated spectra, suggesting that they could be
indiscernible in an experiment. There, other mechanisms
with larger fingerprints are in play and may veil such
negligible shifts, e.g., surface roughness~\cite{pecharroman_prb77,ciraci_acsphot7}, at
least for individual samples. This suggests that indeed,
following the perhaps na\"{i}ve but intuitive expectations,
it is possible to identify situations where the nonlocality
of the system is hidden in mutually canceling contributions.

To clearly demonstrate that nonlocal effects do play a role 
in this system, Fig.~\ref{fig:toymodelPW} also shows the two
cases where both NPs are described by the same-sign Feibelman
parameters. These spectra behave as one would expect from
the response of individual NPs, i.e., resonance redshifts
are observed when spill-out is present (see yellow line)
and the effective interparticle gap decreases, while
blueshifts characterize the case of spill-in dominating
(light blue line). The line of reasoning in the case
of dimers can thus be understood from an intuitive
geometric point of view: spill-out and spill-in can be
seen as effectively altering the dimensions of the
monomers in question and by extension the size of
the gap (potentially leaving it unaltered when the
increase in the first case balances the decrease in
the latter). In the literature of mesoscopic plasmonics,
such a perspective has appeared early
on~\cite{teperik_prl110,teperik_oex21} and has proven a
valuable tool to help understand the impact of
nonlocality in homodimer systems (or at the very least
how it is embedded in popular mesoscopic models). At
the same time, it has also been recognized that simple
geometric rescaling is not sufficient to explain the
richness of the optical response of a mesoscopic
system~\cite{luo_prl111}.

The spectra of Fig.~\ref{fig:toymodelPW} constitute one
characteristic example of the cancellation under study,
which can be observed for a wide range of heterodimer
parameters. To reach cancellation, an interplay between
nonlocality due to the size of the individual NPs and due
to the shrinking of the nanogap needs to be considered.
If the NPs are too small, individual frequency shifts in
each of them might detune their respective dipolar
localized surface plasmons (LSPs), decreasing the
strength of plasmon hybridization once they are brought
close together. For this reason, in our toy-model example
we focus on relatively large particles, where the gap
role shall be more stressed. On the other hand,
depending on the predominant nonlocal response,
spill-in or -out, the effective gap in the dimer might
increase or decrease so much that, again, the hybridization
strength can be considerably affected. The most striking
example in Fig.~\ref{fig:toymodelPW} is the $d^{+}$ spectrum,
where the two combined spill-out corrections effectively
decrease the gap from $3.0$\,nm to $\sim 2.2$\,nm,
leading to dramatic redshifts of the BDPs, and the clear
manifestation of a third-order BDP at $\sim 2.25$\,eV,
almost overlapping with the BDP2 predicted by LRA.
Note here that we deliberately avoid
minimizing the considered gaps further, as to
avoid entering the tunneling regime~\cite{savage_nat491},
for which the SRF framework is not appropriate, and
effective treatments such as the quantum-corrected 
model might be needed~\cite{esteban_natcom3}.

\begin{figure*}[ht]
\centering\includegraphics[width=0.9\linewidth]{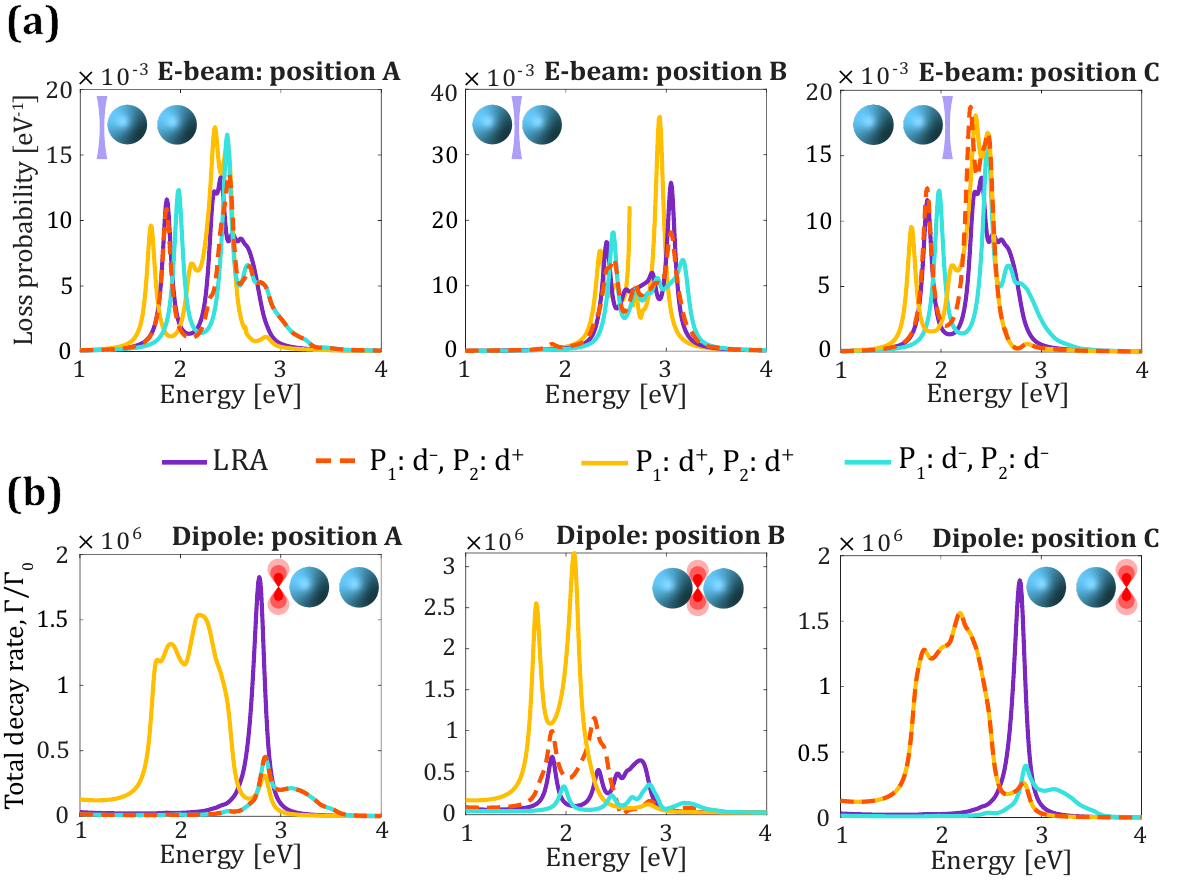}
\caption{(a) EEL spectra for the dimer of Fig.~\ref{fig:toymodelPW},
with the electron beam traveling on the left of the dimer, through
the center of the gap, or on the right of the dimer (positions A,
B, and C, from left to right).
(b) Normalized total decay rate for point dipoles placed at
positions A, B, and C (from left to right), with their dipole
moment along the dimer axis.
In all panels, the colors of the lines follow the convention
introduced in Figs.~\ref{fig:sketch} and \ref{fig:toymodelPW}.
}\label{fig:toymodelLocalized}
\end{figure*}

So far, the response of the heterodimer has been
explored with plane-wave excitation which, in the
far-field, produces the average response of the system.
It is tempting to examine if the effective cancellation
of nonlocal corrections also occurs for a localized
excitation, such as an electron beam or a dipole.
In the past, these approaches has proven valuable in
probing the physical mechanisms governing nonlocality
of single NPs within HDM~\cite{christensen_nn8,zouros_prb101}
or SRF~\cite{goncalves_nl23}. The local~\cite{koh_nn3}
and nonlocal response~\cite{zhang_adqt1} of dimers made
of identical NPs has also been explored with swift electron
beams. Heterodimers have been studied with electron
energy loss spectroscopy (EELS)~\cite{flauraud_nn11},
which allows to excite and characterize all optical modes
of the system. We thus start our discussion in
Fig.~\ref{fig:toymodelLocalized}(a)
with EELS calculations, for electron beams with velocity
$\mathbf{v} \simeq 0.78c \hat{\mathbf{z}}$ passing nearby
the dimer at the three positions sketched in
Fig.~\ref{fig:sketch}:
A, 1\,nm to the left of the leftmost (spill-in) NP;
B, in the middle of the gap, along the dimer axis;
C, 1\,nm to the right of the rightmost (spill-out) NP.

The first thing that one can observe in the figure is that
for excitations on the sides the coincidence between LRA
and SRF for BDP1 is maintained. On the contrary,
higher-order-mode predictions within the two models are
in worse agreement than in the case of plane-wave excitation.
The higher-order radiative and nonradiative modes of
individual NPs are, in general, unevenly influenced by
nonlocality, with modes of higher angular momentum being
subjected to stronger shifts~\cite{christensen_nn8}.
Considering the modes of the dimer as results of
hybridization of higher (and potentially mutually
different)-order modes of individual NPs, the intuitive
cancellation of spill-in and spill-out cannot happen
for the dimensions used in Fig.~\ref{fig:toymodelPW}. There,
parameters were chosen with optical spectroscopy in mind,
and the extinction of the dimer can mainly capture radiative
modes with dipolar origin. Nevertheless, it is interesting
to observe that when the electron beam passes near the
spill-in NP (left), there is a good spectral match between
the full SRF spectrum and the $d^{-}$ case. The same can
also be observed when the electron passes close to the
spill-out NP (right), where the higher-order BDPs now
agree with the $d^{+}$ spectrum. This suggests that
while the nonlocality of the BDP1 mode can still not be
resolved, the electron beam can unveil the influence
of nonlocality on all higher-order modes. However, this
picture is altered when the electron beam passes through
the center of the gap (B), where, again, the full SRF
result matches the LRA prediction: the system ---including
excitation--- is fully symmetric, and thus the two SRF
corrections to the LRA cancel. In this panel, it can also
be observed that BDP1 is not excited in LRA, as expected
for symmetry reasons, since in the case of identical
particles and excitation exactly at the center there is
no way to excite a mode with opposite charges on the
two sides of the gap~\cite{koh_nn3,nelson_quillin_jpcc120}.
However, within SRF the two NPs are no longer identical,
as they are characterized by a small amount of spill-in or
spill-out. This breaking of the symmetry leads to a small
detectable signal, as can be seen by the small resonance
(orange line) at $\sim 1.9$\,eV. Importantly, this spectral
feature constitutes a manifestation of the surface response,
providing a direct signature of the underlying
spill-in/spill-out effects.

With electron beams disclosing a different nonlocal
response at different excitation spots, the natural next
step is to explore point dipole sources, again at different
positions regarding the dimer. This should allow to
probe the LDOS of the system, which has proven to differ
from EELS near the NP~\cite{hohenester_prl103}. Rather than
trying to make a direct connection to the EELS calculations,
here we opt again to study the configuration most relevant
to experiments, namely an electron dipole oscillating along
the dimer axis. As shown in the insets of
Fig.~\ref{fig:toymodelLocalized}(b), we probe the dimer
with point dipoles placed at $s = 1$\,nm from the surface 
of the two NPs, or exactly at the center of the gap.
Interestingly, for this small distance, when the dipole
is on the sides of the dimer, it only ''sees'' its nearest
NP, while the other NP remains in the ''shadow'' of the
first. The normalized modal decay rate $\Gamma/\Gamma_{0}$
in Fig.~\ref{fig:toymodelLocalized}(b)--position A, as
calculated within SRF, is considerably different from the
LRA prediction, but it practically coincides with the $d^{-}$
spectrum. All the energy of the dipole immediately couples
with all radiative and nonradiative multipoles of the
spill-in NP, leading to a small blueshift of the first-order
resonance, and a broader feature related to nonradiative
coupling to higher-order multipoles. We note that convergence
in this energy window is rather challenging and this
finding might not be entirely trustworthy, as multipolar
studies have shown a continuous shift of the broad
feature~\cite{goncalves_natcom11}. Nevertheless, when we
place the dipole at $s = 1$\,nm from the right (spill-out)
NP, the Purcell factor coincides again with the one
predicted for the $d^{+}$ scenario
[Fig.~\ref{fig:toymodelLocalized}(b)--position C]. Again,
we observe significant redshifts, as one would expect for
spill-out, and a considerable increase in the absolute value
of the Purcell factor, related to increased absorptive losses
as the dipole is effectively brought even closer to the
surface of the NP. Finally, placing the dipole at the center
of the gap [Fig.~\ref{fig:toymodelLocalized}(b)--position B]
leads to an intermediate situation, where the Purcell factor
for BDP1 matches well the LRA prediction, but all higher-order
modes become much more lossy. The resonant feature at
$\sim 2.3$\,eV suggests that the SRF prediction might be
emerging as a combination of BDP2 predictions within the
different models, as the dipole probes both the spill-in
and the spill-out NP (see the corresponding spectra in
Fig.~\ref{fig:toymodelPW}). This does make sense, as the
effective distance of the dipole from the surface of the
spill-in NP increases, and that NP plays a smaller role,
but the distance from the spill-out NP decreases and
approaches the example of position C.

\section{A realistic example}

\begin{figure*}[ht]
\centering\includegraphics[width=0.8\linewidth]{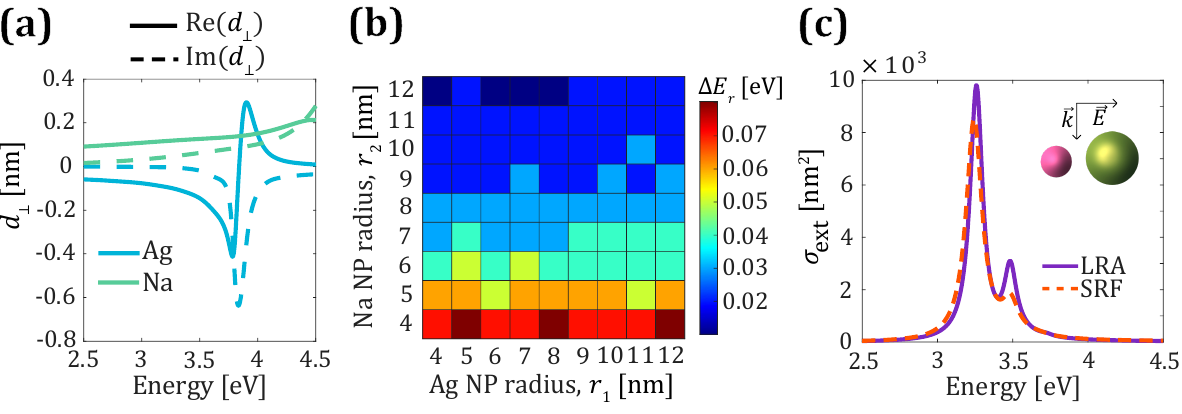}
\caption{(a) Real (solid lines) and imaginary (dashed lines)
part of the Feibelman $d_{\perp}(\omega)$ parameter adopted
here for Ag (light blue lines) and Na (green lines).
(b)~Parametric-space scan for nonlocal cancellation.
The gap between the two NPs is kept constant, $g = 3$\,nm,
while their radii vary from $4$ to $12$\,nm. The color
bar depicts the difference in BDP1 resonance energy
between LRA and SRF.
(c) Extinction cross-section of a Na--Ag heterodimer,
with $r_{1} = 4$\,nm (Ag), $r_{2} = 12$\,nm (Na) and gap
$g = 3$\,nm. The LRA (purple solid line) and SRF (orange
dashed line) predictions are almost identical for the
BDP1 mode, and agree very well in terms of energy for
BDP2 as well.
}\label{fig:AgNaMap}
\end{figure*}

Everything so far was demonstrated for an idealized
system, where both NPs were made of alkali-like
Drude metals, and their Feibelman parameters were
constant. We shall now explore whether this effective
cancellation of nonlocal corrections and their 
subsequent re-emergence with localized sources can happen
for realistic material parameters. To this end, we consider
heterodimers made of Na (as the spill-out example) and Ag
(as the spill-in case). The two metals are described by the
Feibelman $d_{\perp}(\omega)$ parameters shown in
Fig.~\ref{fig:AgNaMap}(a). Due to the difference in plasma
frequencies of the two metals, we need to extend our
search by adding one more degree of freedom, which
we choose to be the NP size, while we maintain the gap
size constant. In doing so, the underlying LRA system 
already lacks the mirror symmetry of our previous case,
something that should also clearly manifest in the
EEL spectra.

In Fig.~\ref{fig:AgNaMap}(b) we present a parametric-space
map for the Ag--Na dimer. Scanning over different radii
for the two NPs, we plot in color the difference in
resonance energies between the LRA and SRF calculations,
$\Delta E_{\mathrm{r}}$, focusing on the BDP1 mode. The
map suggests that the cancellation we are interested in
can occur for a relatively large range of Ag radii
($r_{1}$), as long as the Na NP is larger than
$\sim 10$\,nm. This can be understood by resorting to
the nonlocal response of the individual NPs: in the
quasistatic regime, following the approximate expression
$\omega_{1} = \omega_{\mathrm{p}}/\sqrt{3}$ (where the
plasma frequency might be modified to
$\omega_{\mathrm{p}}/\sqrt{\varepsilon_{\mathrm{b}}}$ if
the background permittivity is different from one) for
the dipolar LSP, the Ag NPs are characterized by a
resonance at about $3.25$\,eV, while Na NPs have their
resonance closer to $3.4$\,eV. Then, the role of
nonlocality is to blueshift the resonances of Ag to
higher energies, while for Na a redshift would be
obtained, pushing the two resonances closer to each
other. For the given gap size, these opposite trends
reach a cancellation for small Ag NPs facing relatively
large Na ones (top left corner of the map). It is also
important to note that, in addition to the role of
nonlocality on each NP, one has to also consider the
different hybridization strengths between LSPs that
shift with size, making the identification of the ideal
example mainly a numerical task. In Fig.~\ref{fig:AgNaMap}(c)
we show one example of canceling spectra, corresponding to
$r_{1} = 4$\,nm for Ag and $r_{2} = 12$\,nm for Na, lying
indeed in the top-left corner of the map. Here, owing
also to the different damping rates that characterize
the two materials, the BDP1 and BDP2 cancellation is
not as perfect as in the toy-model case, but both
resonances of the dimer exhibit very good agreement
between the LRA and SRF cases, and also linewidths that
are very similar ---especially if one considers how they
would look in an experiment, where surface roughness and
measuring uncertainties lead to further broadening. This
shall thus be the example to adopt for the dipole and
electron-beam studies next.

\begin{figure}[htbp]
\centering\includegraphics[width=1.0\columnwidth]{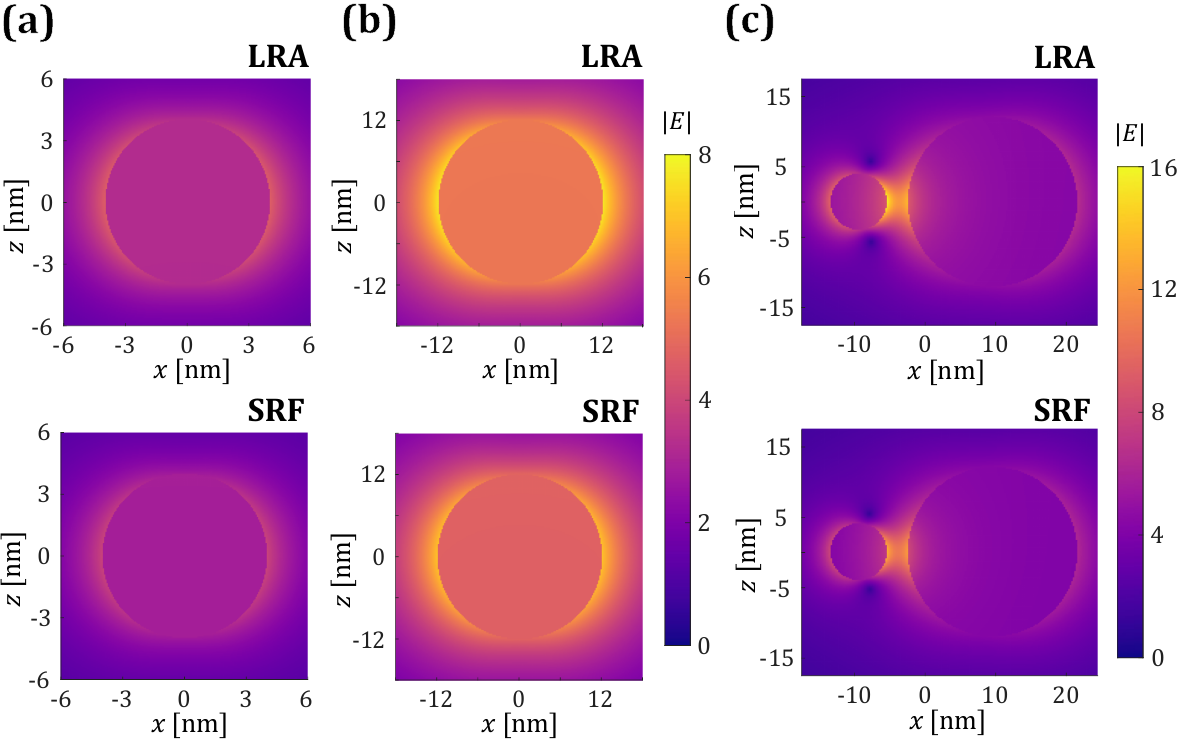}
\caption{Near-field plots showing the magnitude of the electric
field surrounding (a)~a single Ag NP with $r_{1} = 4$\,nm at
$\hbar\omega = 3.49/3.53$\,eV,
(b) a single Na NP with $r_{2} = 12$\,nm at 
$\hbar\omega = 3.35/3.31$\,eV and 
(c) a heterodimer comprised of the two particles,
at $\hbar\omega = 3.26/3.24$\,eV,
The fields are shown in both the LRA (top row) and
SRF (bottom row) frameworks, and the resonant energies given
above follow the order LRA/SRF. The single-NP plots
share one colorbar in both models, while the dimers follow
a second colorbar; both represent the total electric field
normalized to the incident electric field.}
\label{fig:NearFields}
\end{figure}

\begin{figure*}[ht]
\centering\includegraphics[width=0.8\linewidth]{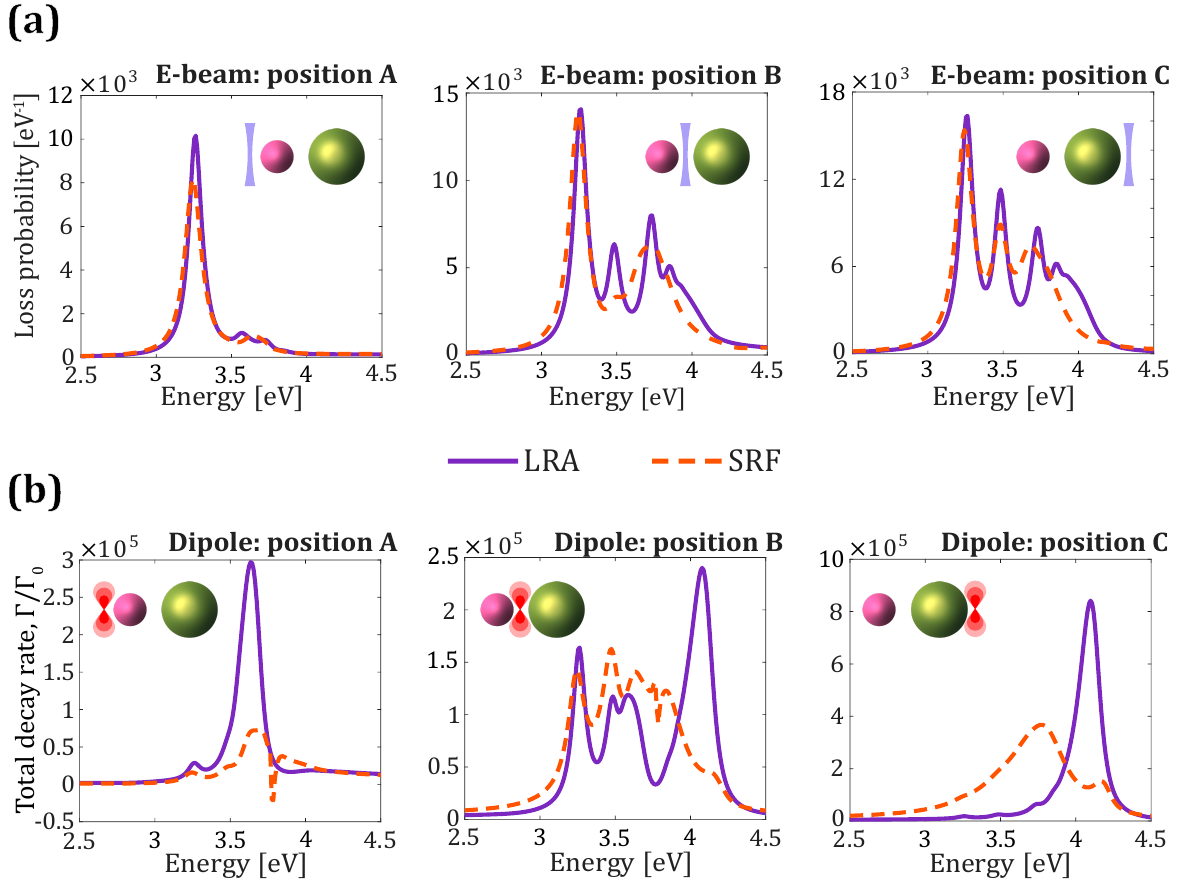}
\caption{(a) Electron-beam excitation of the Ag--Na heterodimer,
for three different positions as depicted in the insets. Purple 
lines depict the total loss probability calculated within LRA,
while orange lines correspond to SRF calculations.
(b) Same as (a), for a point-dipole probing at the same spots,
and the total decay rate. In all cases the dipole moment is along
the dimer axis, as sketched in the insets.
}\label{fig:AgNaLocalized}
\end{figure*}

This discussion about the interplay between the individual-NP
and dimer-hybridization roles of nonlocality is further
supported by the near-field profiles of
Fig.~\ref{fig:NearFields}. The figure shows both the
near-field profiles of the individual NPs on resonance,
as obtained within LRA and SRF [panels (a) and (b)], and
the response of the dimer [panel (c)]. It is evident that
the fields of the individual NPs decay faster away from the
surface within SRF, which reduces the coupling strength at
a given gap size when they are brought together to build
the dimer. At the same time, the resonances of the individual
NPs move farther away within SRF, meaning that a larger
interaction strength is needed~\cite{prodan_jcp120} to form
a hybridized state at the same energy as compared to LRA.
Finally, the spill-out effect in the Na NP extends over
a larger distance than the spill-in in Ag, also affecting
the interaction. The combination of all these trends
directly affects the final NP interaction and the emergence
of the final hybrid resonances, yet which one will prevail
is not immediately obvious. One could be tempted to introduce
a hybridization metric to quantify all these effects, but
this goes beyond the scope of this work.

In Fig.~\ref{fig:AgNaLocalized}(a), the Ag--Na dimer
discussed above is probed by a swift electron beam,
passing again at points A, B, and C as depicted in
Fig.~\ref{fig:sketch}. Compared to the toy model, the
spectra present much smaller differences between LRA
and SRF, regardless of the electron position. Apart
from the additional broadening of higher-order coupled
modes, the energy shifts themselves are experimentally
irrelevant, suggesting that it might be only the point
dipoles that can truly disentangle the different nonlocal
contributions. This can again be anticipated by the
near-field plots of Fig.~\ref{fig:NearFields},
considering how the coupling of the evanescent
electron beam might be affected. We thus turn to
electric point dipoles in Fig.~\ref{fig:AgNaLocalized}(b),
maintaining the same positions and orientations as in Fig.~\ref{fig:toymodelLocalized}(b). Similarly
to that figure, when the dipoles are on
the sides of the dimer, and very close to one of its two
components, the SRF total decay rate changes considerably
from the LRA calculation, resembling more the result of a
dipole close to one NP only. On the other hand, when the
dipole is placed in the center, a much richer spectrum is
observed in both LRA and SRF, as the dipole now sees 
efficiently both NPs and allows their coupling. For
example, focusing on the LRA case, one can indeed
identify the resonance at about $4.1$\,eV originating from
the Na NP, the resonances around $3.5$\,eV related to the
Ag NP, and at $3.3$\,eV the BDP mode, as identified with
plane-wave excitation in Fig.~\ref{fig:AgNaMap}(c). The
latter is not significantly affected by nonlocality, as the
LRA and SRF results practically overlap. It should also be
mentioned that an unrealistic spike also appears in the
spectra at about $3.79$\,eV, for both positions A and B.
One might wonder if this behavior could be cured by enforcing
passivity for the Feibelman parameter of Ag (e.g., setting
the imaginary part positive), but this violates
causality~\cite{svendsen_jpcm32}. However, we have 
verified that these spectra maintain unphysical dips
and as such the culprit must be found elsewhere. The spike
can be directly attributed to the resonant feature in the
Feibelman parameters of Ag at the same energy, and the
proximity of the dipole to the NP. For such small 
dipole--NP separations, the emitter couples to all
higher-order multipoles of the sphere, for which the
wavenumber $k_{\ell}$ scales with multipole-order 
$\ell$, meaning that the local assumption $ks \ll 1$
breaks down. For the real-material example, this
is the only energy window in our work
where local Feibelman parameters prove inaccurate, and one
should resort to the nonlocal
ones~\cite{babaze_oex30,babaze_nanoph12}. Nevertheless, we
stress that both the qualitative discussion and also, to a
great extent, the exact results are not otherwise affected.

\section{Conclusion}

To summarize, we posed the question whether a simple
geometric interpretation of plasmonic nonlocality, and
in particular the spill-in and spill-out effects
characterizing noble and alkali metals, respectively,
can lead to an effective cancellation of nonlocal modal
shifts in the optical spectra of NP heterodimers.
Through a toy-model with simple spill-in/out contributions
described by constant, non-dispersive Feibelman parameters,
we showed that this is indeed feasible. To then retrieve the
nonlocal response of the dimer, we showed that localized 
excitations, such as electron beams and point dipoles
provide flexibility and deeper insight into the plasmonic
coupling. In realistic materials, the cancellation under
study can still emerge if one adds an additional free
parameter, e.g., the NP size or their distance. Once again,
while the nonlocal effects tend to cancel each other
out in the far-field, the near-field response of the
system, particularly the decay rate of nearby dipoles
(and thus the LDOS) prove very sensitive to configuration.
This suggests that small molecules and quantum dots can
act as efficient probes for the nonclassical response of
plasmonic systems. Similarly, quantum corrections to
plasmonic cavities need to be accounted for when
combined with QEs.

\section*{Acknowledgments}
We thank Javier Aizpurua for useful discussions,
and Christian Nicolaisen~Hansen for creative input on the visualizations.
The Center for Polariton-driven Light--Matter Interactions (POLIMA)
is funded by the Danish National Research Foundation (Project No. DNRF165).
P.~E.~S. is a VILLUM International Postdoc supported
by VILLUM FONDEN (VIL71383).
P.~A.~D.~G is a VILLUM Young Investigator supported
by VILLUM FONDEN (VIL71383).
C.~T. acknowledges support from Independent Research Fund
Denmark (5281-00155B).\\

\section*{Author contributions}
C.~T., N.~A.~M., and P.~E.~S. conceived the idea.
L.~H. and U.~H. prepared the BEM toolbox.
P.~E.~S. obtained preliminary results.
P.~E.~S. and C.~R. ran \textsc{treams} simulations
for validation.
F.~W.~C. and P.~A.~D.~G. ran multiple-scattering
simulations and prepared the near-field plots.
I.~J.~B. and C.~M. validated the results, performed
the bulk of the calculations, and analyzed the data
with C.~T, while I.~J.~B. prepared the figures.
C.~T., C.~M., I.~J.~B., and N.~A.~M. wrote the first
draft of the manuscript.
All authors contributed to editing the manuscript and
discussing the results.

\section*{Data availability}
Data underlying the results presented in this paper
are not publicly available at this time but may be
obtained from the authors upon reasonable request.

\bibliography{references.bib}

@article{acimovic_nn3,
author = {A\'{c}imovi\'{c}, S.~S. and Kreuzer, M.~P. and
Gonz\'{a}lez, M.~U. and Quidant, R.},
title = {Plasmon near-field coupling in metal dimers as a
step toward single-molecule sensing},
journal = {ACS Nano},
volume = {3},
number = {4},
pages = {1231--1237},
year = {2009},
doi = {10.1021/nn900102j}
}

@article{anger_prl96,
author = {Anger, P. and Bharadwaj, P. and Novotny, L.},
title = {Enhancement and quenching of single-molecule
fluorescence},
journal = {Phys. Rev. Lett.},
volume = {96},
number = {11},
pages = {113002},
year = {2006},
doi = {10.1103/PhysRevLett.96.113002}
}

@article{babaze_oex30,
author = {Babaze, A. and Ogando, E. and
Stamatopoulou, P.~E. and Tserkezis, C. and
Mortensen, N.~A. and Aizpurua, J. and
Borisov, A.~G. and Esteban, R.},
title = {Quantum surface effects in the electromagnetic
coupling between a quantum emitter and a plasmonic
nanoantenna: time-dependent density functional theory
vs. semiclassical {Feibelman} approach},
journal = {Opt. Express},
volume = {30},
number = {12},
pages = {21159--21183},
year = {2022},
doi = {10.1364/oe.456338}
}

@article{babaze_nanoph12,
author = {Babaze, A. and Neuman, T. and Esteban, R. and
Aizpurua, J. and Borisov, A.~G.},
title = {Dispersive surface-response formalism to address
nonlocality in extreme plasmonic field confinement},
journal = {Nanophotonics},
volume = {12},
number = {16},
pages = {3277--3289},
year = {2023},
doi = {10.1515/nanoph-2023-0178}
}

@article{bundgaard_josab41,
author = {Bundgaard, I.~J. and {Nicolaisen Hansen}, C. and
Stamatopoulou, P.~E. and Tserkezis, C.},
title = {Quantum-informed plasmonics for strong coupling:
the role of electron spill-out},
journal = {J. Opt. Soc. Am. B},
volume = {41},
number = {5},
pages = {1144--1151},
year = {2024},
doi = {10.1364/josab.512129}
}

@article{cha_bkcs36,
author = {Cha, H. and Yoon, S.},
title = {{Ag--Au} nanoparticle dimers: high-yield preparation
and plasmon coupling for various interparticle distances},
journal = {Bull. Kor. Chem. Soc.},
volume = {36},
number = {3},
pages = {1040--1043},
year = {2015},
doi = {10.1002/bkcs.10143}
}

@article{chandra_jacs134,
author = {Chandra, M. and Dowgiallo, A.-M. and Knappenberger~Jr., K.~L.},
title = {Magnetic dipolar interactions in solid gold nanosphere dimers},
journal = {J. Phys. Chem. C},
volume = {134},
number = {10},
pages = {4477-4480},
year = {2012},
doi = {10.1021/ja210648a}
}

@article{chang_prl97,
author = {Chang, D.~E. and S{\o}rensen, A.~S. and 
Hemmer, P.~R. and Lukin, M.~D.},
title = {Quantum optics with surface plasmons},
journal = {Phys. Rev. Lett.},
volume = {97},
number = {5},
pages = {053002},
year = {2006},
doi = {10.1103/PhysRevLett.97.053002}
}

@article{chen_pnas122,
author = {Chen, Z. and Yang, S. and Xie, Z. and
Hu, J. and Zhang, X. and Xia, Y. and Shen, Y. and
Su, H. and Xie, M. and Christensen, T. and Yang, Y.},
title = {Broadband measurement of {Feibelman}’s quantum
surface response functions},
journal = {Proc. Natl. Acad. Sci.},
volume = {122},
number = {23},
pages = {e2501121122},
year = {2025},
doi = {10.1073/pnas.2501121122}
}

@article{christensen_nn8,
author = {Christensen, T. and Yan, W. and Raza, S. and
Jauho, A.-P. and Mortensen, N.~A. and Wubs, M.},
title = {Nonlocal response of metallic nanospheres probed
by light, electrons, and atoms},
journal = {ACS Nano},
volume = {8},
number = {2},
pages = {1745--1758},
year = {2014},
doi = {10.1021/nn406153k}
}

@article{christensen_prl118,
author = {Christensen, T. and Yan, W. and Jauho, A.-P. and
Solja\v{c}i\'{c}, M. and Mortensen, N.~A.},
title = {Quantum corrections in nanoplasmonics: shape, scale,
and material},
journal = {Phys. Rev. Lett.},
volume = {118},
number = {15},
pages = {157402},
year = {2017},
doi = {10.1103/PhysRevLett.118.157402}
}

@article{ciraci_sci337,
author = {Cirac\`{i}, C. and Hill, R.~T. and Mock, J.~J. and
Urzhumov, Y. and Fern\'{a}ndez-Dom\'{i}nguez, A.~I. and
Maier, S.~A. and Pendry, J.~B. and Chilkoti, A. and
Smith, D.~R.},
title = {Probing the ultimate limits of plasmonic enhancement},
journal = {Science},
volume = {337},
number = {6098},
pages = {1072--1074},
year = {2012},
doi = {10.1126/science.1224823}
}

@article{ciraci_prb93,
author = {Cirac\`{i}, C. and {Della Sala}, F.},
title = {Quantum hydrodynamic theory for plasmonics:
impact of the electron density tail},
journal = {Phys. Rev. B},
volume = {93},
number = {20},
pages = {205405},
year = {2016},
doi = {10.1103/PhysRevB.93.205405}
}

@article{ciraci_nanoph8,
author = {Cirac\`{i}, C. and Jurga, R. and Khalid, M. and
{Della Sala}, F.},
title = {Plasmonic quantum effects on single-emitter
strong coupling},
journal = {Nanophotonics},
volume = {8},
number = {10},
pages = {1821--1833},
year = {2019},
doi = {10.1515/nanoph-2019-0199}
}

@article{ciraci_acsphot7,
author = {Cirac\`{i}, C. and Vidal-Codina, F. and
Yoo, D. and Peraire, J. and Oh, S.-H. and Smith, D.~R.},
title = {Impact of surface roughness in nanogap plasmonic
systems},
journal = {ACS Photonics},
volume = {7},
number = {4},
pages = {908--913},
year = {2020},
doi = {10.1021/acsphotonics.0c00099}
}

@article{ding_adpr4,
author = {Ding, T. and Tserkezis, C. and Mystilidis, C. and
Vandenbosch, G.~A.~E. and Zheng, X.},
title = {Quantum mechanics in plasmonic nanocavities: from
theory to applications},
journal = {Adv. Phys. Res.},
volume = {4},
number = {4},
pages = {2400144},
year = {2025},
doi = {10.1002/apxr.202400144}
}

@article{drexhage_bbpcl72,
author = {Drexhage, K.~H. and Kuhn, H. and Sch\"{a}fer, F.~P.},
title = {Variation of the fluorescence decay time of a molecule
in front of a mirror},
journal = {Ber. Bunsenges. Phys. Chem. Lett.},
volume = {72},
issue = {2},
pages = {329},
year = {1968},
doi = {10.1002/bbpc.19680720261}
}

@article{eriksen_nanoph13,
author = {Eriksen, M.~H. and Tserkezis, C. and
Mortensen, N.~A. and Cox, J.~D.},
title = {Nonlocal effects in plasmon-emitter interactions},
journal = {Nanophotonics},
volume = {13},
number = {15},
pages = {2741--2751},
year = {2024},
doi = {10.1515/nanoph-2023-0575},
}

@article{esteban_natcom3,
author = {Esteban, R. and Borisov, A.~G. and Nordlander, P. and
Aizpurua, J.},
title = {Bridging quantum and classical plasmonics with a
quantum-corrected model},
journal = {Nat. Commun.},
volume = {3},
pages = {825},
year = {2012},
doi = {10.1038/ncomms1806}
}

@article{feibelman_pss12,
author = {Feibelman, P.~J.},
title = {Surface electromagnetic fields},
journal = {Prog. Surf. Sci.},
volume = {12},
issue = {4},
pages = {287--508},
year = {1982},
doi = {10.1016/0079-6816(82)90001-6}
}

@article{fernandez_acsphot5,
author = {Fern{\'a}ndez-Dom{\'i}nguez, A.~I. and Bozhevolnyi, S.~I. and
Mortensen, N.~A.},
title = {Plasmon-enhanced generation of nonclassical light},
journal = {ACS Photonics},
volume = {5},
number = {9},
pages = {3447--3451},
year = {2018},
doi = {10.1021/acsphotonics.8b00852}
}

@article{flauraud_nn11,
author = {Flauraud, V. and Bernasconi, G.~D. and Butet, J. and
Alexander, D.~T.~L. and Martin, O.~J.~F. and Brugger, J.},
title = {Mode coupling in plasmonic heterodimers probed with
electron energy loss spectroscopy},
journal = {ACS Nano},
volume = {11},
number = {4},
pages = {3485--3495},
year = {2017},
doi = {10.1021/acsnano.6b08589}
}

@article{ford_physrep113,
author = {Ford, G.~W. and Weber, W.~H.},
title = {Electromagnetic interactions of molecules
with metal surfaces},
journal = {Phys. Rep.},
volume = {113},
number = {4},
pages = {195--287},
year = {1984},
doi = {10.1016/0370-1573(84)90098-X}
}

@article{fu_lpr3,
author = {Fu, Y. and Lakowicz, J.~R.},
title = {Modification of single molecule fluorescence near
metallic nanostructures},
journal = {Laser Photon. Rev.},
volume = {3},
issue = {1--2},
pages = {221--232},
year = {2009},
doi = {10.1002/lpor.200810035}
}

@article{abajo_jpcc112,
author = {Garc\'{i}a de Abajo, F.~J.},
title = {Nonlocal effects in the plasmons of strongly interacting
nanoparticles, dimers, and waveguides},
journal = {J. Phys. Chem. C},
volume = {112},
number = {46},
pages = {17983--17987},
year = {2008},
doi = {10.1021/jp807345h}
}

@article{goncalves_natcom11,
author = {Gon\c{c}alves, P.~A.~D. and Christensen, T. and Rivera, N. and
Jauho, A.-P. and Mortensen, N.~A. and Solja\v{c}i\'{c}, M.},
title = {Plasmon–emitter interactions at the nanoscale},
journal = {Nat. Commun.},
volume = {11},
pages = {366},
year = {2020},
doi = {10.1038/s41467-019-13820-z}
}

@article{goncalves_nl23,
author = {Gon\c{c}alves, P.~A.~D. and Garc\'{i}a de Abajo, F.~J.},
title = {Interrogating quantum nonlocal effects in nanoplasmonics
through electron-beam spectroscopy},
journal = {Nano Lett.},
volume = {23},
number = {10},
pages = {4242--4249},
year = {2023},
doi = {10.1021/acs.nanolett.3c00298}
}

@article{halevi_prb51,
author = {Halevi, P.},
title = {Hydrodynamic model for the degenerate free-electron gas:
generalization to arbitrary frequencies},
journal = {Phys. Rev. B},
volume = {51},
issue = {12},
pages = {7497--7499},
year = {1995},
doi = {10.1103/PhysRevB.51.7497}
}

@article{hohenester_prl103,
author = {Hohenester, U. and Ditlbacher, H. and Krenn, J.~R.},
title = {Electron-energy-loss spectra of plasmonic nanoparticles},
journal = {Phys. Rev. Lett.},
volume = {103},
number = {10},
pages = {106801},
year = {2009},
doi={10.1103/PhysRevLett.103.106801}
}

@article{hohenester_prb105,
author = {Hohenester, U. and Unger, G.},
title = {Nanoscale electromagnetism with the boundary element
method},
journal = {Phys. Rev. B},
volume = {105},
number = {7},
pages = {075428},
year = {2022},
doi = {10.1103/PhysRevB.105.075428}
}

@article{hohenester_comphy276,
author = {Hohenester, U. and Reichelt, N. and Unger, G.},
title = {Nanophotonic resonance modes with the {NANOBEM} toolbox},
journal = {Comput. Phys. Commun.},
volume = {276},
pages = {108337},
year = {2022},
doi = {10.1016/j.cpc.2022.108337}
}

@article{huang_jpcc130,
author = {Huang, Z. and Ge, W.-X. and Zhang, Y. and Gao, L.},
title = {Giant near-field thermal rectification enabled by
nonlocal resonance coupling in asymmetric nanophotonic dimers},
journal = {J. Phys. Chem. C},
volume = {130},
number = {20},
pages = {7051--7060},
year = {2026},
doi = {10.1021/acs.jpcc.6c00562}
}

@article{huber_jpcc129,
author = {Huber, L. and Hohenester, U.},
title = {A computational {Maxwell} solver for nonlocal
{Feibelman} parameters in plasmonics},
journal = {J. Phys. Chem. C},
volume = {129},
number = {5},
pages = {2590--2598},
year = {2025},
doi = {10.1021/acs.jpcc.4c07387}
}

@Article{Uskov_plasmonics9,
  Title = {Broadening of plasmonic resonances due to electron collisions with nanoparticle boundary: quantum-mechanical consideration},
  Author = {Uskov, A. V. and Protsenko, I. E. and Mortensen, N. A. and O'Reilly, E. P.},
  Journal = {Plasmonics},
  Year = {2014},
  Pages = {185-192},
  Volume = {9},
  doi ={10.1007/s11468-013-9611-1}
}

@article{johnson_prb6,
author = {Johnson, P.~B. and Christy, R.~W.},
title = {Optical constants of the noble metals},
journal = {Phys. Rev. B},
volume = {6},
issue = {12},
pages = {4370--4379},
year = {1972}, 
doi = {10.1103/PhysRevB.6.4370}
}

@article{jurga_jpcc121,
author = {Jurga, R. and D'Agostino, S. and Della Sala, F. and
Cirac\`{i}, C.},
title = {Plasmonic nonlocal response effects on dipole decay dynamics
in the weak- and strong-coupling regimes},
journal = {J. Phys. Chem. C},
volume = {121},
issue = {40},
pages = {22361--22368},
year = {2017},
doi = {10.1021/acs.jpcc.7b07462}
}

@article{khurgin_acsphot4,
author = {Khurgin, J.~B and Tsai, W.-Y. and
Tsai, D.~P. and Sun, G.},
title = {Landau damping and limit to field confinement and
enhancement in plasmonic dimers},
journal = {ACS Photonics},
volume = {4},
issue = {11},
pages = {2871--2880},
year = {2017},
doi = {10.1021/acsphotonics.7b00860}
}

@article{koenderink_nl9,
author = {Koenderink, A.~F.},
title = {Plasmon nanoparticle array waveguides for single
photon and single plasmon sources},
journal = {Nano Lett.},
volume = {9},
number = {12},
pages = {4228--4233},
year = {2009},
doi = {10.1021/nl902439n}
}

@article{koenderink_ol35,
author = {Koenderink, A.~F.},
title = {On the use of {Purcell} factors for plasmon
antennas},
journal = {Opt. Lett.},
volume = {35},
number = {24},
pages = {4208--4210},
year = {2012},
doi = {10.1364/OL.35.004208}
}

@article{koh_nn3,
author = {Koh, A.~L. and Bao, K. and Khan, I. and Smith, W.~E. and
Kothleitner, G. and Nordlander, P. and Maier, S.~A. and {McComb}, D.~W.},
title = {Electron energy-loss spectroscopy ({EELS}) of surface plasmons in
single silver nanoparticles and dimers: influence of beam damage and mapping of
dark modes},
journal={ACS Nano},
volume={3},
number={10},
pages={3015--3022},
year={2009},
doi={10.1021/nn900922z}
}

@article{kreibig_zphys224,
author = {Kreibig, U. and {von Fragstein}, C.},
title = {The limitation of electron mean free path in small
silver particles},
journal = {Z. Phys.},
volume = {224},
issue = {4},
pages = {307--323},
year = {1969},
doi = {10.1007/bf01393059}
}

@article{kreibig_surfsci156,
author = {Kreibig, U. and Genzel, L.},
title = {Optical absorption of small metallic particles},
journal = {Surf. Sci.},
volume = {156},
issue = {2},
pages = {678--700},
year = {1985},
doi = {10.1016/0039-6028(85)90239-0}
}

@article{kumar_acsphot8,
author = {Kumar, S and Bozhevolnyi, S.~I.},
title = {Single-photon generation engineering},
journal = {ACS Photonics},
volume = {8},
number = {11},
pages = {3119--3124},
year = {2021},
doi = {10.1021/acsphotonics.1c00849}
}

@article{li_admat36,
author = {Li, Y. and Chen, W. and He, X. and Shi, J. and
Cui, X. and Sun, J. and Xu, H.},
title = {Boosting light--matter interactions in plasmonic
nanogaps},
journal = {Adv. Mater.},
volume = {36},
number = {49},
pages = {2405186},
year = {2024},
doi = {10.1002/adma.202405186}
}

@article{li_njp15,
author = {Li, X. and Xiao, D. and Zhang, Z.},
title = {Landau damping of quantum plasmons in metal nanostructures},
journal = {New J. Phys.},
volume = {15},
issue = {2},
pages = {023011},
year = {2013},
doi = {10.1088/1367-2630/15/2/023011}
}

@article{luo_prl111,
author = {Luo, Y. and Fernandez-Dominguez, A.~I. and
Wiener, A. and Maier, S.~A. and Pendry, J.~B.},
title = {Surface plasmons and nonlocality: a simple model},
journal = {Phys. Rev. Lett.},
volume = {111},
issue = {9},
pages = {093901},
year = {2013},
doi = {10.1103/PhysRevLett.111.093901}
}

@article{martin_prb84,
author = {Mart\'{i}n-Cano, D. and Gonz\'{a}lez-Tudela, A. and
Mart\'{i}n-Moreno, L. and Garc\'{i}a-Vidal, F.~J. and Tejedor, C. and
Moreno, E.},
title = {Dissipation-driven generation of two-qubit entanglement
mediated by plasmonic waveguides},
journal = {Phys. Rev. B},
volume = {84},
issue = {23},
pages = {235306},
year = {2011},
doi = {10.1103/PhysRevB.84.235306}
}

@article{mie_annphys330,
author = {Mie, G.},
title = {Beitr\"{a}ge zur {Optik} tr\"{u}ber {Medien}, speziell 
kolloidaler {Metall\"{o}sungen}},
journal = {Ann. Phys.},
volume = {330},
number = {3},
pages = {377--445},
doi = {10.1002/andp.19083300302},
year = {1908}
}

@article{monticone_omex15,
author = {Monticone, F. and Mortensen, N.~A. and
Fern\'{a}ndez-Dom\'{i}nguez, A.~I. and Luo, Y. and
Zheng, X. and Tserkezis, C. and Khurgin, J.~B. and
Shahbazyan, T.~V. and Chaves, A.~J. and Peres, N.~M.~R. and
Wegner, G. and Busch, K. and Hu, H. and {Della Sala}, F. and
Zhang, P. and Cirac\`{i}, C. and Aizpurua, J. and
Babaze, A. and Borisov, A.~G. and Chen, X.-W. and
Christensen, T. and Yan, W. and Yang, Y. and
Hohenester, U. and Huber, L. and Wubs, M. and
{De Liberato}, S. and Gon\c{c}alves, P.~A.~D. and
{Garc\'{i}a de Abajo}, F.~J. and Hess, O. and
Tarasenko, I. and Cox, J.~D. and Jelver, L. and
Dias, E.~J.~C. and {S\'{a}nchez S\'{a}nchez}, M. and
Margetis, D. and G\'{o}mez-Santos, G. and
Vasilevskiy, I.~M. and Stauber, T. and Tretyakov, S. and
Simovski, C. and Pakniyat, S. and G\'{o}mez-D\'{i}as, J.~S. and
Bondarev, I.~V. and Biehs, S.-A. and Boltasseva, A. and
Shalaev, V.~M. and Krasavin, A.~V. and Zayats, A.~V. and
Al\`{u}, A. and Song, J.-H. and Brongersma, M.~L. and
Levy, U. and Long, O.~Y. and Guo, C. and Fan, S. and
Bozhevolnyi, S.~I. and Overvig, A. and
Prud\^{e}ncio, F.~R. and Silveirinha, M.~G. and
Gangaraj, S.~A.~H. and Argyropoulos, C. and
Huidobro, P.~A. and Galiffi, E. and Yang, F. and
Pendry, J.~B. and Miller, D.~A.~B.},
title = {Nonlocality in photonic materials and metamaterials:
roadmap},
journal = {Opt. Mater. Express},
volume = {15},
number = {7},
pages = {1544--1709},
year = {2025},
doi = {10.1364/OME.559374}
}

@article{mortensen_nanoph10a,
author = {Mortensen, N.~A.},
title = {Mesoscopic electrodynamics at metal surfaces},
journal = {Nanophotonics},
volume = {10},
issue = {10},
pages = {2563--2616},
year = {2021},
doi = {10.1515/nanoph-2021-0156}
}

@article{mortensen_nanoph10b,
author = {Mortensen, N.~A. and Gon\c{c}alves, P.~A.~D. and
Shuklin, F.~A. and Cox, J.~D. and Tserkezis, C. and Ichikawa, M. and
Wolff, C.},
title = {Surface-response functions obtained from equilibrium
electron-density profiles},
journal = {Nanophotonics},
volume = {10},
issue = {14},
pages = {3647--3657},
year = {2021},
doi = {10.1515/nanoph-2021-0084}
}

@article{nelson_quillin_jpcc120,
author = {Nelson-Quillin, A.~M. and Cherqui, C. and
Montoni, N.~P. and Li, G. and Camden, J.~P. and Masiello, D.~J.},
title = {Imaging plasmon hybridization in metal nanoparticle
aggregates with electron energy-loss spectroscopy},
journal = {J. Phys. Chem. C},
volume = {120},
number = {37},
pages = {20852--20859},
year = {2016},
doi = {10.1021/acs.jpcc.6b02170}
}

@article{nordlander_nl4,
author = {Nordlander, P. and Oubre, C. and Prodan, E. and
Li, K. and Stockman, M.~I.},
title = {Plasmon hybridization in nanoparticle dimers},
journal = {Nano Lett.},
volume = {4},
number = {5},
pages = {899--903},
year = {2004},
doi = {10.1021/nl049681c}
}

@article{pecharroman_prb77,
author = {Pecharrom\'{a}n, C. and P\'{e}rez-Juste, J. and
Mata-Osoro, G. and Liz-Marz\'{a}n, L.~M. and Mulvaney, P.},
title = {Redshift of surface plasmon modes of small gold
rods due to their atomic roughness and end-cap geometry},
journal = {Phys. Rev. B},
volume = {77},
number = {3},
pages = {035418},
year = {2008},
doi = {10.1103/PhysRevB.77.035418}
}

@article{prodan_sci302,
author = {Prodan, E. and Radloff, C. and Halas, N.~J. and
Nordlander, P.},
title = {A hybridization model for the plasmon response of
complex nanostructures},
journal = {Science},
volume = {302},
number = {5644},
pages = {419--422},
year = {2003},
doi = {10.1126/science.1089171}
}

@article{prodan_jcp120,
author = {Prodan, E. and Nordlander, P.},
title = {Plasmon hybridization in spherical nanoparticles},
journal = {J. Chem. Phys.},
volume = {120},
number = {11},
pages = {5444--5454},
year = {2004},
doi = {10.1063/1.1647518}
}

@article{raza_jpcm27,
author = {Raza, S. and Bozhevolnyi, S.~I. and Wubs, M. and
Mortensen, N.~A.},
title = {Nonlocal optical response in metallic nanostructures},
journal = {J. Phys.: Condens. Matter},
volume = {27},
issue = {18},
pages = {183204},
year = {2015},
doi = {10.1088/0953-8984/27/18/183204}
}

@article{rodriguez_optica8,
author = {Rodr\'{i}guez Echarri, A. and Gon\c{c}alves, P.~A.~D. and
Tserkezis, C. and {Garc\'{i}a de Abajo}, F.~J. and Mortensen, N.~A. and
Cox, J.~D.},
title = {Optical response of noble metal nanostructures: quantum surface
effects in crystallographic facets},
journal = {Optica},
volume = {8},
number = {6},
pages = {710--721},
year = {2021},
doi = {10.1364/optica.412122}
}

@article{romero_oex14,
author = {Romero, I. and Aizpurua, J. and Bryant, G.~W. and
Garc\'{i}a~de~Abajo, F.~J.},
title = {Plasmons in nearly touching metallic nanoparticles: singular
response in the limit of touching dimers},
journal = {Opt. Express},
volume = {14},
issue = {21},
pages = {9988--9999},
year = {2006},
doi = {10.1364/oe.14.009988}
}

@article{ruppin_prl31,
author = {Ruppin, R.},
title = {Optical properties of a plasma sphere},
journal = {Phys. Rev. Lett.},
volume = {31},
issue = {24},
pages = {1434--1437},
year = {1973},
doi = {10.1103/PhysRevLett.31.1434}
}

@article{savage_nat491,
author = {Savage, K.~J. and Hawkeye, M.~M. and Esteban, R. and
Borisov, A.~G. and Aizpurua, J. and Baumberg, J.~J.},
title = {Revealing the quantum regime in tunnelling plasmonics},
journal = {Nature},
volume = {491},
issue = {7423},
pages = {574--577},
year = {2012},
doi = {10.1038/nature11653}
}

@article{sheikholeslami_nl10,
author = {Sheikholeslami, S. and Jun, Y.-W. and
Jain, P.~K. and Alivisatos, A.~P.},
title = {Coupling of optical resonances in a compositionally
asymmetric plasmonic nanoparticle dimer},
journal = {Nano Lett.},
volume = {10},
issue = {7},
pages = {2655--2660},
year = {2010},
doi = {10.1021/nl101380f}
}

@article{stamatopoulou_omex12,
author = {Stamatopoulou, P.~E. and Tserkezis, C.},
title = {Finite-size and quantum effects in plasmonics: manifestations and
theoretical modelling},
journal = {Opt. Mater. Express},
volume = {12},
number = {5},
pages = {1869--1893},
year = {2022},
doi = {10.1364/ome.456407}
}

@article{stella_jpcc117,
author = {Stella, L. and Zhang, P. and Garc\'{i}a-Vidal, F.~J. and
Rubio, A. and Garc\'{i}a-Gonz\'{a}lez, P.},
title = {Performance of nonlocal optics when applied to plasmonic
nanostructures},
journal = {J. Phys. Chem. C},
volume = {117},
issue = {17},
pages = {8941--8949},
year = {2013},
doi = {10.1021/jp401887y}
}

@article{su_acsabm4,
author = {Su, Q. and Jiang, C. and Gou, D. and Long, Y.},
title = {Surface plasmon-assisted fluorescence enhancing and
quenching: from theory to application},
journal = {ACS Appl. Bio Mater.},
volume = {4},
number = {6},
pages = {4684--4705},
year = {2021},
doi = {10.1021/acsabm.1c00320}
}

@article{svendsen_jpcm32,
author = {Svendsen, M. K. and Wolff, C. and Jauho, A.-P. and
Mortensen, N.~A. and Tserkezis, C.},
title = {Role of diffusive surface scattering in nonlocal plasmonics},
journal = {J. Phys.: Condens. Matter},
volume = {32},
issue = {39},
pages = {395702},
year = {2020},
doi = {10.1088/1361-648X/ab977d}
}

@article{teperik_prl110,
author = {Teperik, T.~V. and Nordlander, P. and Aizpurua, J. and
Borisov, A.~G.},
title = {Robust subnanometric plasmon ruler by rescaling of the
nonlocal optical response},
journal = {Phys. Rev. Lett.},
volume = {110},
number = {26},
pages = {263901},
year = {2013},
doi = {10.1103/PhysRevLett.110.263901}
}

@article{teperik_oex21,
author = {Teperik, T.~V. and Nordlander, P. and Aizpurua, J. and
Borisov, A.~G.},
title = {Quantum effects and nonlocality in strongly coupled
plasmonic nanowire dimers},
journal = {Opt. Express},
volume = {21},
number = {22},
pages = {27306--27325},
year = {2013},
doi = {10.1364/oe.21.027306}
}

@article{toscano_oex20,
author = {Toscano, G. and Raza, S. and Jauho, A.-P. and
Mortensen, N.~A. and Wubs, M.},
title = {Modified field enhancement and extinction by plasmonic
nanowire dimers due to nonlocal response},
journal = {Opt. Express},
volume = {20},
issue = {4},
pages = {4176--4188},
year = {2012},
doi = {10.1364/oe.20.004176}
}

@article{toscano_natcom6,
author = {Toscano, G. and Straubel, J. and Kwiatkowski, A. and
Rockstuhl, C. and Evers, F. and Xu, H. and Mortensen, N.~A. and
Wubs, M.},
title = {Resonance shifts and spill-out effects in self-consistent
hydrodynamic nanoplasmonics},
journal = {Nat. Commun.},
volume = {6},
pages = {7132},
year = {2015},
doi = {10.1038/ncomms8132}
}

@article{tserkezis_nscale8,
author = {Tserkezis, C. and Stefanou, N. and Wubs, M. and
Mortensen, N.~A.},
title = {Molecular fluorescence enhancement in plasmonic
environments: exploring the role of nonlocal effects},
journal = {Nanoscale},
volume = {8},
issue = {40},
pages = {17532--17541},
year = {2016},
doi = {10.1039/c6nr06393d}
}

@article{tserkezis_ijmpb7,
author = {Tserkezis, C. and Yan, W. and Hsieh, W. and Sun, G. and
Khurgin, J.~B. and Wubs, M. and Mortensen, N.~A.},
title = {On the origin of nonlocal damping in plasmonic monomers
and dimers},
journal = {Int. J. Mod. Phys. B},
volume = {31},
issue = {24},
pages = {1740005},
year = {2017},
doi = {10.1142/S0217979217400057}
}

@article{tserkezis_prb96,
author = {Tserkezis, C. and Mortensen, N.~A. and Wubs, M.},
title = {How nonlocal damping reduces plasmon-enhanced fluorescence
in ultranarrow gaps},
journal = {Phys. Rev. B},
volume = {96},
issue = {8},
pages = {085413},
year = {2017},
doi = {10.1103/PhysRevB.96.085413}
}

@article{tserkezis_acsphot5a,
author = {Tserkezis, C. and Wubs, M. and Mortensen, N.~A.},
title = {Robustness of the {Rabi} splitting under nonlocal
corrections in plexcitonics},
journal = {ACS Photonics},
volume = {5},
issue = {1},
pages = {133--142},
year = {2018},
doi = {10.1021/acsphotonics.7b00538}
}

@article{vielma_jcp126,
author = {Vielma, J. and Leung, P.~T.},
title = {Nonlocal optical effects on the fluorescence and decay
rates for admolecules at a metallic nanoparticle},
journal = {J. Chem. Phys.},
volume = {126},
number = {19},
pages = {194704},
year = {2007},
doi = {10.1063/1.2734549}
}

@article{warnakula_prb100,
author = {Warnakula, T. and Gunapala, S.~D. and
Stockman, M.~I. and Premaratne, M.},
title = {Cavity quantum electrodynamic analysis of spasing
in nanospherical dimers},
journal = {Phys. Rev. B},
volume = {100},
number = {8},
pages = {085439},
year = {2019},
doi = {10.1103/PhysRevB.100.085439}
}

@article{wegner_prb107,
author = {Wegner, G. and Huynh, D.-N. and Mortensen, N.~A. and
Intravaia, F. and Busch, K.},
title = {Halevi's extension of the {Euler}--{Drude} model for plasmonic systems},
journal = {Phys. Rev. B},
volume = {107},
number = {11},
pages = {115425},
year = {2023},
doi = {10.1103/PhysRevB.107.115425}
}

@article{xu_prl93,
author = {Xu, H. and Wang, X.-H. and Persson, M.~P. and
Xu, H.~Q. and K\"{a}ll, M. and Johansson, P.},
title = {Unified treatment of fluorescence and
{Raman} scattering processes near metal surfaces},
journal = {Phys. Rev. Lett.},
volume = {94},
number = {24},
pages = {243002},
year = {2004},
doi = {10.1103/PhysRevLett.93.243002}
}

@article{yan_prl115,
author = {Yan, W. and Wubs, M. and Mortensen, N.~A.},
title = {Projected dipole model for quantum plasmonics},
journal = {Phys. Rev. Lett.},
volume = {115},
issue = {13},
pages = {137403},
year = {2015},
doi = {10.1103/PhysRevLett.115.137403}
}

@article{yan_prb91,
author = {Yan, W.},
title = {Hydrodynamic theory for quantum plasmonics:
linear-response dynamics of the inhomogeneous electron gas},
journal = {Phys. Rev. B},
volume = {91},
number = {11},
pages = {115416},
year = {2015},
doi = {10.1103/PhysRevB.91.115416}
}

@article{yang_nat576,
author = {Yang, Y. and Zhu, D. and Yan, W. and Agarwal, A. and
Zheng, M. and Joannopoulos, J.~D. and Lalanne, P. and Christensen, T. and
Berggren, K.~K. and Solja\v{c}i\'{c}, M.},
title = {A general theoretical and experimental framework for nanoscale
electromagnetism},
journal = {Nature},
volume = {576},
issue = {7786},
pages = {248--252},
year = {2019},
doi = {10.1038/s41586-019-1803-1}
}

@article{zhang_adqt1,
author = {Zhang, Q. and Cai, X. and Yu, X. and
Carregal-Romero, S. and Parak, W.~J. and Sachan, R. and
Cai. Y. and Wang, N. and Zhu, Y. and Lei, D.~W.},
title = {Electron energy-loss spectroscopy of spatial nonlocality
and quantum tunneling effects in the bright and dark plasmon modes
of gold nanosphere dimers},
journal = {Adv. Quant. Techn.},
volume = {1},
number = {1},
pages = {1800016},
year = {2018},
doi = {10.1002/qute.201800016}
}

@article{zhu_natcom7,
author = {Zhu, W. and Esteban, R. and Borisov, A.~G. and
Baumberg, J.~J. and Nordlander, P. and Lezec, H.~J. and
Aizpurua, J. and Crozier, K.~B.},
title = {Quantum mechanical effects in plasmonic structures
with subnanometre gaps},
journal = {Nat. Commun.},
volume = {7},
pages = {11495},
year = {2016},
doi = {10.1038/ncomms11495}
}

@article{zouros_prb101,
author = {Zouros, G.~P. and Kolezas, G.~D. and
Mortensen, N.~A. and Tserkezis, C.},
title = {Monitoring strong coupling in nonlocal plasmonics
with electron spectroscopies},
journal = {Phys. Rev. B},
volume = {101},
number = {8},
pages = {085416},
year = {2020},
doi = {10.1103/PhysRevB.101.085416}
}

@article{zuloaga_nl9,
author = {Zuloaga, J. and Prodan, E. and Nordlander, P.},
title = {Quantum description of the plasmon resonances of
a nanoparticle dimer},
journal = {Nano Lett.},
volume = {9},
issue = {2},
pages = {887--891},
year = {2009},
doi = {10.1021/nl803811g}
}

@article{Hohenester2024,
title = {Nanophotonic resonators in stratified media with the nanobem toolbox},
journal = {Computer Physics Communications},
volume = {294},
pages = {108949},
year = {2024},
issn = {0010-4655},
doi = {https://doi.org/10.1016/j.cpc.2023.108949},
url = {https://www.sciencedirect.com/science/article/pii/S0010465523002941},
author = {Ulrich Hohenester}
}

@article{zheng_arxiv2026,
      title={{OpenMUSTANC} {(MUltiple Scattering Theory At Nanoplasmonic Cavities)}: A {MATLAB} toolbox for the simulation of Plasmonic Sphere Aggregates}, 
      author={Zheng, X. and Mystilidis, C. and Tserkezis, C. and Vandenbosch, G. A. E. and Zheng, X.},
      journal = {arXiv:2608.30565},
      doi = {10.48550/arXiv.2608.30565}
}

@article{Ciraci2013,
author = {Cristian Cirac\`{i} and Yaroslav Urzhumov and David R. Smith},
journal = {Opt. Express},
number = {8},
pages = {9397--9406},
publisher = {Optica Publishing Group},
title = {Far-field analysis of axially symmetric three-dimensional directional cloaks},
volume = {21},
month = {Apr},
year = {2013},
url = {https://opg.optica.org/oe/abstract.cfm?URI=oe-21-8-9397},
doi = {10.1364/OE.21.009397},
}

@article{Zheng2026ieee,
  author={Zheng, Xin and Mystilidis, Christos and Tserkezis, Christos and Vandenbosch, Guy A. E. and Zheng, Xuezhi},
  journal={IEEE Trans. Antennas Propag.}, 
  title={An {S}-matrix Formalism for the Nonclassical Optical Response of Plasmonic Sphere Aggregates}, 
  year={2026},
  volume={},
  number={},
  pages={},
  doi={10.1109/TAP.2026.3698275},
  note = {{DOI}: 10.1109/TAP.2026.3698275}
}

@article{Beutel2024,
title = {treams – a {T}-matrix-based scattering code for nanophotonics},
journal = {Comput. Phys. Commun.},
volume = {297},
pages = {109076},
year = {2024},
issn = {0010-4655},
doi = {https://doi.org/10.1016/j.cpc.2023.109076},
url = {https://www.sciencedirect.com/science/article/pii/S0010465523004216},
author = {Dominik Beutel and Ivan Fernandez-Corbaton and Carsten Rockstuhl}
}

@article{Apell1982,
doi = {10.1088/0031-8949/26/2/010},
url = {https://doi.org/10.1088/0031-8949/26/2/010},
year = {1982},
month = {aug},
publisher = {},
volume = {26},
number = {2},
pages = {113},
author = {P. Apell and A. Ljungbert},
title = {A General Non-Local Theory for the Electromagnetic Response of a Small Metal Particle},
journal = {Phys. Scr.}
}
\end{document}